\documentclass[a4paper,11pt,showpacs,amsmath,nofootinbib,superscriptaddress]{revtex4}
\usepackage{graphicx}
\usepackage{comment}
\usepackage{array,amsmath,amsthm,feynmf}
\usepackage{bm}
\usepackage{slashed}
\usepackage{times}
\usepackage{epsfig}
\usepackage{graphicx}
\usepackage{color}
\usepackage{cancel}
\usepackage{amssymb}
\usepackage{textcomp}
\usepackage{mathrsfs}
\newcommand{\be}{\begin{equation}}
\newcommand{\ee}{\end{equation}}
\newcommand{\bea}{\begin{eqnarray}}
\newcommand{\eea}{\end{eqnarray}}

\begin{document}



\title{\Large  Electroweak baryogenesis  in the framework of  the effective field theory }

\vspace{4.0cm}
\author{Fa Peng Huang}
\affiliation{School of Physics and State Key
Laboratory of Nuclear Physics and Technology, Peking
University, Beijing, 100871, China}

\author{Chong Sheng Li\footnote{csli@pku.edu.cn}}
\affiliation{School of Physics and State Key
Laboratory of Nuclear Physics and Technology, Peking
University, Beijing, 100871, China}
\affiliation{Center for High Energy Physics, Peking
University, Beijing, 100871, China}
\begin{abstract}
We study the electroweak baryogenesis in the framework of the effective field theory. Our study shows that by introducing
a light singlet scalar particle and a
dimension-5 operator,  it can provide the strong first order phase transition and the source of  the $CP$-violation during the
phase transition, and then produce abundant particle phenomenology at zero temperature. We also show the constraints on the new physics scale from the
observed baryon-to-photon ratio, the low-energy experiments and the
LHC data.
\end{abstract}
\maketitle
\section{Introduction}\label{sec:Introduction}

The discovery of the Higgs boson at the LHC opens the door for studying the scalar sector of the standard model (SM), and
exploring  the  structure of the  scalar sector will
be an important task for the LHC in the coming years, which can
help us to understand the true mechanism of the electroweak phase transition and
 the origin of baryon asymmetry of  the  universe (BAU).
The BAU, which has been a long unsolved problem in cosmology and particle physics,
is quantified by the baryon-to-photon ratio $\eta=n_B/n_{\gamma}=6.05(7)\times 10 ^{-10}$(\text{CMB)\cite{Ade:2013zuv,Agashe:2014kda}},
where $n_B$ and $n_\gamma$ are the baryon and photon densities, respectively.
The observed value of baryon-to-photon ratio can be  determined from studies of the  power spectrum of the cosmic microwave background radiation
or the big bang nucleosynthesis.
At the end of the inflationary epoch,  to generate the
BAU (baryogenesis) proposed  by Sakarov \cite{Sakharov:1967dj}, three conditions must be satisfied:  baryon number violation, $C$ and $CP$-violation (CPV), and departure from thermal equilibrium or CPT violation.

To solve the baryogenesis problem,  several mechanisms [i.e. Planck-scale baryogenesis, GUT baryogenesis, leptogenesis, Affine-Dine baryogenesis and electroweak baryogenesis (EWB)]
 have been proposed \cite{Dine:2003ax}, but after
the discovery of the $125$ GeV scalar boson \cite{Aad:2012tfa,Chatrchyan:2012ufa},
EWB \cite{Kuzmin:1985mm,Trodden:1998ym} became a popular and testable scenario for explaining the BAU \cite{Morrissey:2012db}.
An important ingredient for the success  of EWB
is the existence of a strong first order phase transition (SFOPT).
However, the $125$ GeV Higgs boson is too heavy for efficient SFOPT \cite{Morrissey:2012db},
and there exist three types of extensions of the SM scalar sector to resolve the inefficiency~\cite{Chung:2012vg}.
Another important ingredient is enough CPV  source, since the CPV source is
too weak in the SM.

In this paper, in the framework of the effective field theory (EFT)(we follow the effective Lagrangian approaches to investigate the EWB in Refs.\cite{Zhang:1993vh,Zhang:1994fb}),
 we introduce a light scalar particle S and an interesting  dimension-5 operator $y_t \frac{\eta}{\Lambda} S \bar{Q}_L \tilde{\Phi} t_R+\rm H.c.$
to provide both the SFOPT and  enough CPV for EWB.
During the SFOPT($\langle S \rangle=\sigma$)\footnote{ In this paper, the angle brackets denote the vacuum expectation value of the field.}, this dimension-5 operator can provide the CPV source for BAU; then
after the SFOPT ($\langle S \rangle=0$), this operator can naturally avoid the strong constraints from the data of electric dipole moments (EDM) 
and yield distinctive signals at the LHC, i.e. monojet plus missing transverse energy (MET), mono-Higgs plus MET, and $\bar{t}t$ plus MET. Meanwhile, we will give the constraints on the parameters of the effective Lagrangian from cosmology and particle physics experiments.

In Sec. \ref{sec:Model}, we describe the effective Lagrangian, which can explain the baryogenesis and produce abundant particle phenomenology.
In Sec. \ref{sec:EWPT}, we discuss the realization of the SFOPT in detail, including the constraints from Higgs invisible decay.
In Sec. \ref{sec:EB},  constraints from the observed baryon-to-photon ratio are obtained.
In Sec. \ref{sec:EDM}, the constraints from the EDM on the new physics (NP) scale are given.
In Sec. \ref{sec:monojet}, we investigate the collider constraints on the NP scale at the LHC.
Finally, we conclude in Sec. \ref{sec:conclusion}.

\section{ The Effective  Lagrangian}\label{sec:Model}


Instead of discussing the baryogenesis in a concrete UV-complete theory
(such as supersymmetric baryogenesis), which is not easy to make confident
predictions about  since it has large sets of undetermined
additional parameters, we try  
to explain the BAU and discuss the possible collider signals at the LHC, using  EFT.
For example, recent studies using the EFT techniques
have considered the NP signals at colliders such as monojet, monophoton and mono-Higgs,  recoiling
against some MET at colliders.
In this paper, we will consider the collider signals of the EWB from the effective Lagrangian:
\begin{eqnarray}
{\cal L}&=&{\cal L}_{\rm SM}+\frac{1}{2} \partial_\mu S \partial^\mu S  + \frac{1}{2}\mu^2  S^2
- \frac{1}{4}\lambda S^4  - \frac{1}{2} \kappa S^2 (\Phi^\dagger \Phi) \label{olag} \\
&+& y_t \frac{\eta}{\Lambda} S \bar{Q}_L \tilde{\Phi} t_R+\rm H.c.   \nonumber
,
\end{eqnarray}
where $\eta=a+i b$ is a complex parameter,  $y_t=\sqrt{2} m_t/v$ is the SM top Yukawa coupling, $\Lambda$ is the NP scale,
$S$ is a light  singlet scalar particle beyond the SM,
$\Phi$ is the  SM Higgs doublet field, $Q_L$ is the $SU(2)_L$ quark doublet, and $t_R$ is the right-handed
top quark. $\lambda$ and $\lambda_{SM}$ are assumed to  be positive  here.
The similar Lagrangian has been investigated  in Refs.~\cite{Espinosa:2011eu,Cline:2012hg}, where
the collider phenomenology has not been discussed.~


For the effective Lagrangian given in Eq.(\ref{olag}),
SFOPT will occur when the vacuum transitions  from  $(0,\langle S \rangle)$ to $(\langle \Phi \rangle,0)$, which
will be discussed  in the following.
During the SFOPT, the scalar field S acquires the vacuum expectation value (VEV) as  $\langle S \rangle$ and
the dimension-5 operator can be rewritten as $\frac{ y_t \langle S \rangle}{\sqrt{2} \Lambda}(a H \bar{t} t+ib H \bar{t}\gamma_5 t)$.
Thus,
the top quark mass gets a
spatially varying complex phase along the bubble wall profile~\cite{Espinosa:2011eu,Cline:2012hg},
which provides the source of CPV needed to generate the BAU.

At zero temperature, the VEV of S vanishes and the dimension-5 operator
can avoid the  electric EDM constraints and induce the interaction term $\frac{ m_t}{\Lambda}(a S \bar{t} t+ib S \bar{t}\gamma_5 t)$,
which would produce abundant collider signals, such as
monojet plus MET, mono-Higgs plus MET, and $\bar{t}t$ plus MET at the LHC.

\section{SFOPT}\label{sec:EWPT}
\subsection{Vacuum structure at  tree level and zero temperature}


Since the phase transition is largely influenced by the vacuum property of the scalar sector at zero temperature,
we first study the vacuum structure at tree level and zero temperature.
If only the vacuum  is considered,
we can write the potential as a function of a singlet VEV $\langle S(x) \rangle= \sigma(x)$ 
and the Higgs field VEV $\langle \Phi(x) \rangle=\frac{1}{\sqrt{2}} (0, H(x))^T$,
and we can simplify  the potential  as
\begin{equation}\label{treev}
V_\text{tree}(H,\sigma)=-\frac{1}{2}\mu_{SM}^2 H^2-\frac{1}{2}\mu^2 \sigma^2+\frac{1}{4}\lambda_{SM} H^4+\frac{1}{4}\lambda \sigma^4+\frac{1}{4}\kappa H^2\sigma^2.
\end{equation}
The extremal points can be obtained by the minimization conditions:
\begin{equation}\label{min}
\frac{\partial V_\text{tree}}{\partial H}\Big|_{(H,\sigma)}=\frac{\partial V_\text{tree}}{\partial \sigma}\Big|_{(H,\sigma)}=0\,.
\end{equation}
Among the nine extremal points, there  exist four distinct extremal points as
\begin{eqnarray}
   (H,\sigma)&=& (0,0), \\
 (H,\sigma) &=& (\frac{\mu_{SM}}{\sqrt{\lambda_{SM}}},0), \\
  (H,\sigma) &=& (0,\frac{\mu}{\sqrt{\lambda}}), \\
  (H,\sigma) &=& \left(\sqrt{\frac{4 \lambda\mu_{\rm SM}^2 - 2 \kappa \mu^2 }
   { 4 \lambda \lambda_{\rm SM} - \kappa^2 }},\sqrt{ \frac{4 \lambda_{\rm SM}\mu^2 -2 \kappa \mu_{\rm SM}^2 }
 { 4 \lambda \lambda_{\rm SM} - \kappa^2 }}\right),
\end{eqnarray}
The corresponding effective potentials are
\begin{eqnarray}
  V(0,0)&=& 0 ,\\
  V(\frac{\mu_{SM}}{\sqrt{\lambda_{SM}}},0) &=&  -\frac{\mu_{SM}^{4}}{4 \lambda_{SM}} ,\\
  V(0,\frac{\mu}{\sqrt{\lambda}})&=&  -\frac{\mu^{4}}{4 \lambda}, \\
  V\left(\sqrt{\frac{4 \lambda\mu_{\rm SM}^2 - 2 \kappa \mu^2 }
   { 4 \lambda \lambda_{\rm SM} - \kappa^2 }},\sqrt{ \frac{4 \lambda_{\rm SM}\mu^2 -2 \kappa \mu_{\rm SM}^2 }
 { 4 \lambda \lambda_{\rm SM} - \kappa^2 }}\right) &=& \frac{\lambda_{SM}\mu^4+\lambda \mu_{SM}^4-\mu^2 \mu_{SM}^2 \kappa}{\kappa^2-4 \lambda \lambda_{SM}}.
\end{eqnarray}
Since $\lambda$ and $\lambda_{SM}$ are assumed to  be positive, then $V(\frac{\mu_{SM}}{\sqrt{\lambda_{SM}}},0) < 0$ and $ V(0,\frac{\mu}{\sqrt{\lambda}})<0$.
For simplicity, we further need  $ V(0,\frac{\mu}{\sqrt{\lambda}})$ and $V(\frac{\mu_{SM}}{\sqrt{\lambda_{SM}}},0)$ to be the global minimum, namely, $ V\left(\sqrt{\frac{4 \lambda\mu_{\rm SM}^2 - 2 \kappa \mu^2 }
   { 4 \lambda \lambda_{\rm SM} - \kappa^2 }},\sqrt{ \frac{4 \lambda_{\rm SM}\mu^2 -2 \kappa \mu_{\rm SM}^2 }
 { 4 \lambda \lambda_{\rm SM} - \kappa^2 }}\right)>V(\frac{\mu_{SM}}{\sqrt{\lambda_{SM}}},0)$ and
 $ V\left(\sqrt{\frac{4 \lambda\mu_{\rm SM}^2 - 2 \kappa \mu^2 }
   { 4 \lambda \lambda_{\rm SM} - \kappa^2 }},\sqrt{ \frac{4 \lambda_{\rm SM}\mu^2 -2 \kappa \mu_{\rm SM}^2 }
 { 4 \lambda \lambda_{\rm SM} - \kappa^2 }}\right)> V(0,\frac{\mu}{\sqrt{\lambda}})$.
 These requirements  lead to
 \begin{equation}\label{kkk}
 \kappa>2 \sqrt{\lambda \lambda_{SM}}.
 \end{equation}
The degenerate condition of the two minima at tree level is
\begin{equation}\label{deg_cond}
\frac{\mu_{SM}^{4}}{ \lambda_{SM}}=\frac{\mu^{4}}{ \lambda},
\end{equation}
which is useful for future discussion of the SFOPT.~If $\frac{\mu_{SM}^{4}}{ \lambda_{SM}}>\frac{\mu^{4}}{ \lambda}$,
then $V(\frac{\mu_{SM}}{\sqrt{\lambda_{SM}}},0)$ is the only global minimum.

\subsection{Loop and thermal effects }
Following the methods in Refs. \cite{Quiros:1999jp,Dolan:1973qd}, the full finite-temperature
effective potential up to one-loop level is composed of three parts,
\begin{equation}\label{fullpotential}
 V_{eff}(H,\sigma,T)=V_\text{tree}(H,\sigma)+
V_1^{T=0}(H,\sigma)+\Delta V_1^{T\neq 0}(H,\sigma,T),
\end{equation}
where $V_\text{tree}(H,\sigma)$ is the tree-level potential in Eq.(\ref{treev}) ; $V_1^{T=0}(H,\sigma)$ is the Coleman-Weinberg
potential at zero temperature; and $\Delta V_1^{T\neq 0}(H,\sigma,T)$ is the leading thermal correction.
Using the high-temperature expansion up to $\mathcal{O}(T^2)$,  the effective thermal potential
 Eq.(\ref{fullpotential}) can
be written as
\begin{equation}\label{eq:highT_V}
 V(H,\sigma;T)=D_H(T^2-T_{0H}^2)H^2+D_\sigma(T^2-T_{0\sigma}^2)\sigma^2
+\frac{1}{4}(\lambda_{SM} H^4+ \kappa H^2\sigma^2 + \lambda \sigma^4),
\end{equation}
with
\begin{align*}
D_H&=\frac{1}{32}(8\lambda_{SM}+g'^2+3 g^2+4 y_t^2+2\kappa),\\
D_\sigma&=\frac{1}{24}(2\kappa+5 \lambda+6g_2^2),\\
T_{0H}^2&=\frac{\mu^2_{SM}}{2D_H},\\
T_{0\sigma}^2&=\frac{\mu^2}{2D_\sigma}\,.
\end{align*}
The values of the SM couplings $g'$, $g$, $y_t$ and $g_2$ are chosen according to the
results in Ref. \cite{Buttazzo:2013uya}.
The terms $D_H T^2$ and $D_{\sigma} T^2$ correspond to the thermal corrections to the mass of H
and $\sigma$ particle, respectively. Here, we omit the thermal contribution of the dimension-5
operator as in Refs.~\cite{Espinosa:2011eu,Cline:2012hg}.

After including the thermal mass effects, the minima of the effective potential become
\begin{eqnarray}
V_{eff}(0,0)&=& 0 ,\\
  V_{eff}(\sqrt{\frac{\mu_{SM}^2-D_H T^2}{\lambda_{SM}}},0,T) &=&  -\frac{(\mu_{SM}^{2}-D_H T^2)^2}{4 \lambda_{SM}} ,\\
  V_{eff}(0,\sqrt{\frac{\mu^2-D_{\sigma} T^2}{\lambda}},T)&=&  -\frac{(\mu^{2}-D_{\sigma} T^2)^2}{4 \lambda},
\end{eqnarray}

\begin{eqnarray}
 V_{eff}\left(\sqrt{\frac{4 \lambda\mu_{\rm SM}^2 - 2 \kappa \mu^2 -4 \lambda D_H T^2+2 D_{\sigma} \kappa T^2}
   { 4 \lambda \lambda_{\rm SM} - \kappa^2 }},\sqrt{ \frac{4 \lambda_{\rm SM}\mu^2 -2 \kappa \mu_{\rm SM}^2 +2 \kappa T^2 D_H-4 D_{\sigma}T^2 \lambda_{SM}}
 { 4 \lambda \lambda_{\rm SM} - \kappa^2 }}\right)                                                \\ \nonumber
  =\frac{T^4 \lambda D_H^2+(\mu^2-D_{\sigma}T^2)^2\lambda_{SM}+T^2D_H(\kappa(\mu^2-D_{\sigma}T^2)-2\lambda \mu_{SM}^2)+\mu_{SM}^2(D_{\sigma}\kappa T^2-\kappa \mu^2+\lambda \mu_{SM}^2)}{\kappa^2-4 \lambda \lambda_{SM}}.
\end{eqnarray}

Due to the vacuum structure, the phase transitions take place through two steps:
first, S acquires a VEV $\langle S \rangle$, and second, $\langle S \rangle$ vanishes as H acquires a VEV $\langle \Phi \rangle$; i.e.
the phase transitions take place as $(0,0) \to (0,\langle S \rangle) \to (\langle \Phi \rangle,0)$ with the decreasing of the
temperature, and SFOPT will occur during the second step from $(0,\langle S \rangle)$ to $(\langle \Phi \rangle,0)$.

Since the phase transition is dominantly controlled by the tree-level scalar potential, we expect the SFOPT
may take place in the vicinity of the degenerate point in Eq.(\ref{deg_cond}). Also, we consider the thermal effects with the small perturbation
($\mid \delta_{\mu^2} \mid \ll 1$ and $\mid \delta_\lambda \mid \ll 1$) around the degenerate point, which is given by
\begin{eqnarray}
  \mu^2 &=& \mu^2_{SM}\frac{\kappa}{2\lambda_{SM}} (1+\delta_{\mu^2}), \label{v1} \\
  \lambda &=&(\frac{\kappa}{2 \lambda_{SM}})^2 \lambda_{SM}(1+\delta_\lambda).\label{v2}
\end{eqnarray}
The critical values can be obtained by substituting Eqs.(\ref{v1}) and (\ref{v2}) into the following expression:
\begin{equation}\label{aa}
V_{eff}(H_c^{(1)},\sigma_c^{(1)},T_c)=V_{eff}(H_c^{(2)},\sigma_c^{(2)},T_c),
\end{equation}
and  the phase transition critical temperature is given by
\begin{equation}\label{tc}
T_c\approx\frac{m_H \sqrt{\delta_\lambda-2\delta_{\mu^2}}}{2 \sqrt{D_h-D_{\sigma}}}.
\end{equation}
From Eqs.(\ref{aa}) and (\ref{tc}), the washout parameter can be obtained as
\begin{equation}\label{vtc}
\frac{v(T_c)}{T_c}\approx  \frac{2  v \sqrt{D_H-D_{\sigma}}}{m_H \sqrt{\delta_\lambda-2\delta_{\mu^2}}}.
\end{equation}
The necessary condition of SFOPT  for baryogenesis is
\begin{equation}
\frac{v(T_c)}{T_c}>1.
\end{equation}
\begin{figure}
\begin{center}
\includegraphics[height=0.3 \textwidth]{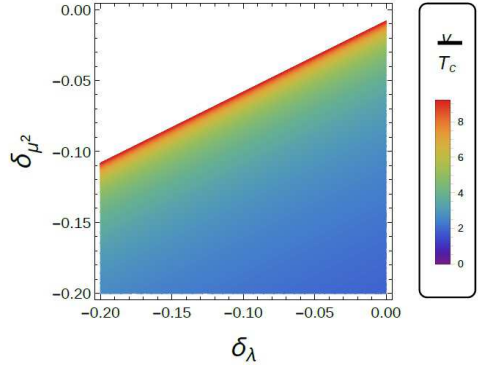}
\caption{The contour plot of the washout parameter $v/T_c$ in the $(\delta_\lambda,\delta_{\mu^2})$ plane.}
\label{wash}
\end{center}
\end{figure}
From Eq.(\ref{vtc}), we see that if $\delta_\lambda-2\delta_{\mu^2} \ll 1$, $v(T_c)/T_c>1$
(such as $\delta_\lambda=\delta_{\mu^2}=-0.1$),  then the SFOPT can be realized. The dependence of the washout parameter $\frac{v(T_c)}{T_c}$ on
$\delta_\lambda$ and $\delta_{\mu^2}$ is shown in Fig. \ref{wash}.
From Eq.(\ref{aa}), we can get
\begin{equation}\label{phtr}
V_{eff}(\sigma,0,T)-V_{eff}(0,H,T) \approx \frac{\mu^4_{SM}}{4 \lambda_{SM}}(\delta_\lambda-2\delta_{\mu^2})-\frac{\mu^2_{SM}}{2 \lambda_{SM}}(D_H-D_{\sigma}) T^2.
\end{equation}
With the decreasing of the temperature,  the global minimum of the effective potential changes from $(\sigma,0)$ to $(0,H)$ and
the SFOPT occurs, which can be seen from Eq.(\ref{phtr}).

At the critical temperature, there exists a barrier between $(0,\langle S \rangle)$ and  $ (\langle \Phi \rangle,0)$, and this leads to
the condition $\kappa > 2 \sqrt{\lambda \lambda_{SM}}$ at $T_c$.
Substituting Eq.(\ref{v2}) and $\mid \delta_\lambda \mid \ll 1$  into this condition, we can get $\kappa > \kappa \sqrt{1+\delta_{\lambda}}$, which gives
$-1 \ll \delta_{\lambda}<0$.
$\delta_{\mu^2}$ is also a small negative value to guarantee the positive value
of $\delta_\lambda-2\delta_{\mu^2}$. Note that this constraint of $\kappa > 2 \sqrt{\lambda \lambda_{SM}}$ at the critical temperature $T_c$
is ignored in Ref.~\cite{Chung:2012vg}\footnote{However, the above constraint is respected in Ref.~\cite{Espinosa:2011eu}.}.
The mass  of the S particle is expressed as
\begin{equation}\label{smass}
m_S^2=-\frac{\kappa v^2 \delta_{\mu^2}}{2}.
\end{equation}
Since we  consider the phase transition with small perturbation
($\mid \delta_{\mu^2} \mid \ll 1$ and $\mid \delta_\lambda \mid \ll 1$) around the
degenerated point, $\mid \delta_{\mu^2} \mid$ and $\mid \delta_\lambda \mid$
should be much smaller than
$1$ as shown in Eqs.(\ref{v1}) and  (\ref{v2}). From the viewpoint of the perturbation theory, $\kappa$ also should
be smaller than $1$, and thus $\mid \kappa \delta_{\mu^2} \mid \ll 1$. From  Eq.(\ref{smass}) for the mass of
the S particle, the above perturbative requirements for $\delta_{\mu^2}$ and $\kappa$  favor a light particle \footnote{The phase transition considered here is similar to the
$\overline{\rm E_c S P'}$ case, which favors a light mass ~\cite{Chung:2012vg}.}(For example, if $\delta_{\mu^2}=-0.125$ and $\kappa=0.6$, then $m_S=47$~GeV), which allows the
Higgs invisible decay~\cite{Chung:2012vg}. The case for the heavy mass has been discussed in
Refs.~\cite{Espinosa:2011eu,Cline:2012hg}, and we only study the light scalar case  with
mass much less than $125$ GeV.
In this scenario, the portal coupling $\kappa$ between the
singlet S and the Higgs boson can be very small as long as
 $\delta_{\lambda}$ and $\delta_{\mu^2}$ are small negative values, so that it can
produce the SFOPT and
satisfy the constraints from Higgs invisible decay below.

\subsection{Higgs invisible decay}\label{sec:Decay}
\begin{figure}[h]
\begin{center}
  \includegraphics[height=0.3 \textwidth]{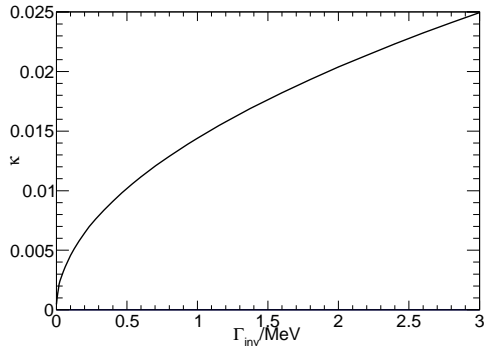}
  \caption{Upper bounds on the Higgs portal coupling from the Higgs invisible
decay width.}\label{invisible}
\end{center}
\end{figure}

After the SFOPT, the VEV of the S field vanishes, and
the SM Higgs doublet field $\Phi$ can be expanded around  the VEV as
$ \Phi(x) =\frac{1}{\sqrt{2}} (0,\langle \Phi(x) \rangle+ H(x))^T$.
Substituting this into the Higgs portal term $ - \frac{1}{2} \kappa S^2 (\Phi^\dagger \Phi)$
in Eq.(\ref{olag}),
we obtain the following interaction term
\begin{equation}
\mathcal{L}_{H \to SS}=-\frac{\kappa \langle \Phi \rangle S^2H}{4},\label{ala}
\end{equation}
which leads to the Higgs invisible decay, and  its
decay width is
\begin{equation}
 \Gamma_{inv}(\rm H \to SS)=\frac{\kappa^2 \langle \Phi \rangle^2}{32 \pi m_H}\sqrt{1-\frac{4m_S^2}{m_H^2}}\simeq\frac{\kappa^2 \langle \Phi \rangle^2}{32 \pi m_H}.\label{widthinv}
\end{equation}
Figure \ref{invisible} shows the relation between the Higgs portal coupling $\kappa$ and $\Gamma_{inv}(H)$.
If we take the  global fit upper bound of the invisible decay width as
\cite{Cheung:2013kla,Cheung:2013oya}
\begin{equation}
 \Gamma_{inv}(\rm H)<1.2 ~ \rm MeV,\label{hi}
\end{equation}
the Higgs portal coupling is constrained as $\kappa<0.016$ from Eqs.(\ref{widthinv})and (\ref{hi}).
This constraint indicates that the mass of the S particle should be lighter than $22$~GeV from Eq.(\ref{smass}).

\section{Constraints from the baryon-to-photon ratio}\label{sec:EB}
\begin{figure}[ht]
\begin{center}
\includegraphics[width=0.5\textwidth,clip]{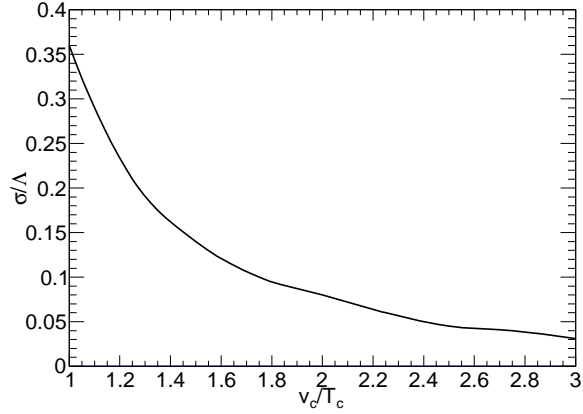}\\
\caption{ \small The approximated numerical relation of the NP scale and the  sphaleron washout parameter during
the SFOPT to produce the observed baryon-to-photon ratio
$\eta=n_B/n_{\gamma}=6.05 \times 10 ^{-10}$ with $L_{\sigma}v_c=5$ and $b=1$. }\label{bau}
\end{center}
\end{figure}

The BAU depends upon a source of CPV that biases
sphaleron interactions near the expanding bubble walls toward
baryon production, as opposed to antibaryons~\cite{Espinosa:2011eu,Cline:2012hg}.
Then, inside the
bubble walls during the SFOPT, the top quark has a
spatially varying complex mass, which is given by~\cite{Espinosa:2011eu,Cline:2012hg}
\begin{equation}
	m_t(z) = {y_t\over\sqrt{2}} H(z) \left(1 +(a+ ib) {S(z)\over\Lambda}\right)\equiv
		|m_t(z)| e^{i\Theta(z)},
\end{equation}
where $z$ is taken to be the coordinate transverse to the wall.
The CPV phase $\Theta$ will provide the source for the BAU, which depends on the sphaleron washout
parameter $v_c/T_c$, the change in VEV $\sigma$ of the singlet, and the
bubble wall thickness $L_\sigma$.
Using the approximated method in Refs.~\cite{Espinosa:2011eu,Cline:2012hg}, the numerical results can be obtained as shown in Fig. \ref{bau},  where
the baryon-to-photon ratio is defined as
\begin{equation}
	\eta_B = {405\Gamma_{\rm sph}\over 4\pi^2v_{\sigma} g_*T}\int dz\, \mu_{B_L}
	f_{\rm sph}\,e^{-45\, \Gamma_{\rm sph}|z|/(4 v_{\sigma})},
\label{baueq}
\end{equation}
and the bubble wall velocity $v_{\sigma}$~\cite{Kozaczuk:2015owa} is chosen as $0.1$ and $\Gamma_{\rm sph} \approx 10^{-6} \rm{T}$.

From the preliminary numerical estimation in Fig.~\ref{bau}, we see that the observed BAU can be obtained
as long as $\sigma/\Lambda < 0.35$.
Since the exact calculation of $\eta_B$ would need improvements of the nonperturbative dynamics,
we will discuss how to  constrain  the NP scale $\Lambda$ from the EDM data and the LHC data below,
which may be more accurate.

\section{Constraints from the neutron EDM}\label{sec:EDM}

Low-energy CPV probes, such as EDMs, lead to
severe constraints on
many baryogenesis models.
For example, the ACME Collaboration's new result, i.e. $\left | \frac{d_e}{e} \right | < 8.7 \times 10^{-29} \, {\rm cm}$ at 90\% confidence level (C.L.) limit
~\cite{Baron:2013eja},
has ruled out  a large  portion of the parameter space for many baryogenesis models.
However, in the case of the considering model in this paper,
the strong constraints from the recent electron EDM experiments can be
naturally avoided.
Due to  the fact that $S$ does not acquire a VEV at zero temperature,
the mixing of $S$ and the Higgs boson and the CPV interaction of Higgs-top is prevented;
i.e. there are no two-loop Barr-Zee contributions to the electric EDM.
Therefore, the dimension-5 operator of the top quark cannot contribute to the electric EDM at the two-loop level.
\begin{figure}[h]
\begin{center}
\includegraphics[height=0.28 \textwidth]{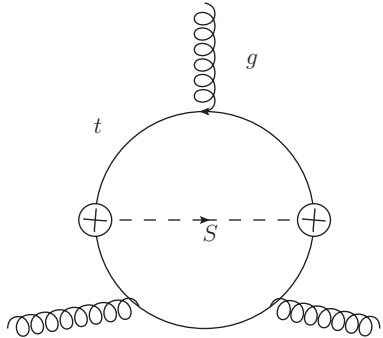}
\caption{The contribution to the Weinberg operator at the two-loop level.}\label{wo}
\end{center}
\end{figure}

However, because of the CPV interaction $\frac{ m_t}{\Lambda}(a S \bar{t} t+ib S \bar{t}\gamma_5 t)$ between the singlet S and top quark at zero temperature,
it can induce the Weinberg operator at the two-loop level, which further gives contribution to
the neutron EDM.
The effective Lagrangian is
\begin{equation}
{\cal L}_W=\frac{-w}{3} f^{abc}G^a_{\mu \sigma} G^{b,\sigma}_{\nu} \tilde{G}^{c,\mu \nu},
\end{equation}
and the corresponding Feynman diagram is  shown in Fig.\ref{wo}.
The two-loop matching coefficient $w$ can be expressed as ~\cite{Dicus:1989va,Braaten:1990gq,Chang:1990jv}
\begin{equation}
w(\mu_W)=\frac{g_s}{4}\frac{\alpha_s}{(4\pi)^3}\sqrt{2}G_F \frac{ab v^2}{\Lambda^2}f_3(x_{t/S}).
\end{equation}
Since the singlet S is very light, $f_3(x_{t/S}) \approx 1$.
After performing numerical calculation,
the contribution 
to the neutron EDM is given by
\begin{equation}
\frac{d_n}{e}=(22\pm 10)\times 2.1 \times 10^{-2} \times \frac{a b v^2}{\Lambda^2} \times 10^{-25}~cm.
\end{equation}
The  $90\%$ C.L. experimental upper bound on the neutron EDM \cite{Baker:2006ts} is
\begin{equation}
\mid \frac{d_n}{e}\mid < 2.9 \times 10^{-26}~cm,
\end{equation}
and higher sensitivity is expected from future experiments \cite{Ito:2007xd}.
After combining the numerical prediction and the experimental bound,
we obtain the constraints on the NP scale $\Lambda$:
\begin{equation}
\Lambda >[229,374]\sqrt{ab}~\rm{GeV}.
\end{equation}
If we choose $a=1$, $b=1$, then  $\Lambda >[229,374]~\rm{GeV}$.
From the above discussion, we see that only if it satisfies $a\neq 0$ and $b \neq 0$ simultaneously does a  contribution to the neutron EDM exist. If $a$ or $b$ becomes zero, we can also avoid the constraints from the neutron
EDM experiments.

\section{Constraints from the monojet plus MET at the LHC}\label{sec:monojet}

\begin{figure}
\begin{center}
  \includegraphics[height=0.15 \textwidth]{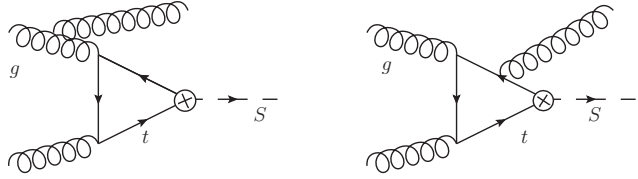}
  \caption{Sample Feynman diagrams for the monojet plus MET signal.}\label{sg}
\end{center}
\end{figure}

At zero temperature, the S field has no VEV, and the Higgs field has a VEV of $v$.
Thus, the dimension-5 operator can induce the operator  $\frac{ m_t}{\Lambda}(ib S \bar{t}\gamma_5 t+ a S \bar{t}\gamma_5 t)$, which
can produce the collider signals
of mono-Higgs plus MET, top pair plus MET and
monojet plus MET. Considering the current experimental precision for these signals, only
the mono-jet plus MET channel is discussed in this paper to give the precise constraints on
the NP scale, since the monojet plus MET channel is the most clean
signal among these channels. The other two channels are beyond this paper's scope,
and we leave 
them for a future work.
The sample Feynman diagrams  for the monojet plus MET signal  are shown in Fig. \ref{sg},
where the S is considered as the MET in collision.
The dominant irreducible SM background for monojet plus MET is $Z+j$
production with $Z$  sequentially decaying to neutrino pairs.

In our numerical calculations, we use
the recent $19.7 fb^{-1}$
of $8$ TeV CMS results \cite{Khachatryan:2014rra}, and  reconstruct jets using the
anti-kt algorithm with radius parameter $R = 0.5$. 
PYTHIA \cite{Sjostrand:2007gs} is used to obtain the parton shower effects.
The  measurements  are performed in seven different
MET regions at CMS, and in our case, we find that for the considered interactions
the highest sensitivity can be obtained for $\rm{MET}>500~\rm{GeV}$.
The cross section for the monojet plus MET signal at $95\%$ C.L. is given by
\begin{equation}\label{section}
\sigma(pp \to \rm MET + jet) < 6.1~\rm{fb},
\end{equation}
\begin{table}[h!]
\centering
\caption{Sample results of the 95\% C.L. lower limits on the NP scale $\Lambda$
from the CMS analysis \cite{Khachatryan:2014rra}. }
\begin{tabular}{ccccc}
\hline
\hline
 $m_{S}$ ~(GeV)  &              ~~~~~~~~~$\Lambda$~(GeV) for $a=b=1$ at $8$ TeV LHC \cite{Khachatryan:2014rra} \\
\hline
6        &  820        \\
12       &  500    \\
\hline
\hline
\end{tabular}

\label{cutoff}
\end{table}
and the constraints on the NP scale $\Lambda$  can be obtained, which are summarized in the Table \ref{cutoff}. We see that
the lower limits of the NP scale are about several hundred GeV from current monojet plus MET data.
Compared to the constraints from the baryon-to-photon ratio and the EDM, the collider experiments provide more strict constraints
on the NP scale.

\section{Conclusions}\label{sec:conclusion}
The discovery of the $125$ GeV scalar particle at the LHC makes the EWB scenario
much more realistic and interesting.
In this paper, we have investigated the phenomenology of the EWG using EFT by
introducing a light scalar particle
and a dimension-5 operator.
We find that the light scalar  field can
give SFOPT as long as $\delta_\lambda$ and $\delta_{\mu^2}$ are small negative values;
the dimension-5 operator can provide the CPV
source to produce the observed baryon-to-photon ratio during the SFOPT and
abundant particle phenomenology.
We also discuss the constraints on the NP
scale from EDM and LHC data, and show
that the extension on the SM with a
light scalar particle and a dimension-5 operator can explain the baryogenesis
problem in the parameter space  allowed by the current EDM and LHC data.

\begin{acknowledgements}
This work was  supported by the National Natural
Science Foundation of China, under Grants No. 11375013 and No. 11135003.
\end{acknowledgements}

\bibliography{BAU_HUANG_NOTE.bib}

\end{document}